\documentstyle[pra,aps,twoside]{revtex}

\newcommand{\ba}{\begin{eqnarray}}
\newcommand{\ea}{\end{eqnarray}}

\begin{document}
\title{Protecting Quantum Information Encoded in Decoherence Free
States Against Exchange Errors}
\author{Daniel A. Lidar$^{(1)}$, David Bacon$^{(1,2)}$, Julia Kempe$^{(1,3)}$ and K. Birgitta Whaley$^{(1)}$}
\address{Chemistry$^{(1)}$, Physics$^{(2)}$, and Mathematics$^{(3)}$
Departments\\
The University of California, Berkeley, CA 94720.}
\maketitle

\begin{abstract}
The exchange interaction between identical qubits in a quantum information
processor gives rise to unitary two-qubit errors. It is shown here that
decoherence free subspaces (DFSs) for collective decoherence undergo Pauli
errors under exchange, which however do not take the decoherence free states
outside of the DFS. In order to protect DFSs against these errors it is
sufficient to employ a recently proposed concatenated DFS-quantum error
correcting code scheme [D.A. Lidar, D. Bacon and K.B. Whaley, Phys. Rev.
Lett. {\bf 82}, 4556 (1999)].\newline

\end{abstract}

\section{Introduction}

Preserving the coherence of quantum states and controlling their unitary
evolution is one of the fundamental goals of Quantum Information Processing 
\cite{Lo:book}. When the system Hamiltonian is invariant under particle
permutations, the exchange operator $E_{ij}$ interchanging particles $i$ and 
$j$ is a constant of the motion, and definite symmetry of a state will be
conserved. Models of quantum computers based on identical bosons or fermions
must of course respect this elementary requirement. It was pointed out in a
recent paper \cite{Ruskai:99} that active quantum error correcting codes
(QECCs) \cite{Shor:95Steane:96a} designed to correct independent
single-qubit errors, will fail for {\em identical} particles in the presence
of exchange errors. The reason is that exchange acts as a {\em two}-qubit
error which has the same effect as a simultaneous bit flip on two different
qubits. Of course, QECCs dealing explicitly with multiple-qubit errors are
also available, so that exchange errors can readily be dealt with provided
one accepts longer codewords than are needed to deal with single-qubit
errors \cite{Gottesman:97Knill:97b}. For example, in Ref. \cite{Ruskai:99} a
nine-qubit code is presented which can correct all single-qubit errors and
all Pauli exchange errors. This is to be compared with the five-qubit
``perfect'' code which protects (only) against all single-qubit errors \cite
{Laflamme:96}.
While the nine-qubit code is longer than the ``perfect'' code, it is
shorter than a code required to protect against {\em all} two-qubit errors.

A different error model which has been considered by several authors is that
in which qubits undergo {\em collective}, rather than independent errors 
\cite{Palma:96Duan:98Zanardi:98a,Zanardi:97c,Lidar:PRL98,Lidar:PRL99}. The
underlying physics of this model has a rich history: it dates back at least
to Dicke's quantum optics work on superradiance of atoms coupled to a
radiation field, where it arose in the consideration of systems confined to
a region whose linear dimensions are small compared to the shortest
wavelength of the field \cite{Dicke:54}. The model was later treated
extensively by Agarwal in the context of spontaneous emission \cite
{Agarwal:book}. It was only recently realized, however, that in the
collective decoherence model there exist large decoherence-free subspaces
(DFSs), which are ``quiet'' Hilbert subspaces in which no
environmentally-induced errors occur at all \cite{Zanardi:97c,Lidar:PRL98}.
Such subspaces offer a passive protection against decoherence. Collective
decoherence is an assumption about the manner in which the environment
couples to the system: instead of independent errors, as assumed in the
active QECC approach, one assumes that errors are strongly correlated, in
the sense that all qubits can be permuted without affecting the coupling
between system and bath. This is clearly a very strong assumption, and it
may not hold exactly in a realistic system-bath coupling scenario. To deal
with this limitation, we have shown recently how DFSs can be stabilized in
the presence of errors that perturb the exact permutation symmetry, by
concatenating DFSs with QECCs \cite{Lidar:PRL99}. Concatenation is a general
technique that is useful for achieving fault tolerant quantum computation 
\cite{Knill:98Preskill:99,Aharonov:99}, and trades stability of quantum
information for the price of longer codewords. It is our purpose here to
analyze the effect of exchange errors on DFSs for collective decoherence.
These errors are fundamentally different from those induced by the
system-bath coupling, since they originate {\em entirely} from the internal
system Hamiltonian. We will show that by use of the very same concatenation
scheme as introduced in Ref. \cite{Lidar:PRL99} (which was designed
originally to deal with system-bath induced errors), a DFS can be stabilized
in the presence of exchange errors as well.

The structure of the paper is as follows. We begin by briefly recalling the
origin of the exchange interaction in Sec. \ref{exchange} and present some
Hamiltonians modelling this interaction. We then present, in Sec. \ref
{DFS-review}, a short review of the Hamiltonian theory of DFSs. Next we
discuss in Sec. \ref{const} the simplest model, of constant exchange matrix
elements, and show that DFSs are immune to exchange errors in this case. Our
main result is then presented in Sec. \ref{arbitrary}, when we analyze the
effect of exchange errors in the case of arbitrary exchange matrix elements.
We show that a DFS is invariant under such errors, and conclude that
concatenation with a QECC can generally stabilize DFSs against exchange.

\section{Modelling Exchange in Qubit Arrays}

\label{exchange}

The exchange interaction arises by virtue of permutation symmetry between
identical particles, in addition to some interaction potential. Exchange is
caused by the system Hamiltonian, and is unrelated to the coupling to an
external environment. Exchange thus induces an extraneous {\em unitary}
evolution on the system, but does not lead to decoherence. To model exchange
it is sufficient to consider a Hamiltonian of the form 
\begin{equation}
H_{{\rm ex}}=\frac{1}{2}\sum_{i\neq j}^{K}J_{ij}E_{ij},  \label{eq:Hex}
\end{equation}
where the sum is over all qubit pairs, $J_{ij}$ are appropriate matrix
elements, and 
\begin{equation}
E_{ij}|\epsilon _{1},...,\epsilon _{i},...,\epsilon _{j},...,\epsilon
_{K}\rangle =|\epsilon _{1},...,\epsilon _{j},...,\epsilon _{i},...,\epsilon
_{K}\rangle .  \label{eq:Eij}
\end{equation}
$E_{ij}$ thus written is a general exchange operator operating on qubits $i$
and $j$ of a $K$-qubit state.

Typical examples of Hamiltonian leading to exchange are \cite{March:85vol1}:
(i) the Heisenberg interaction between spins 
\begin{equation}
H_{{\rm Heis}}=\frac{1}{2}\sum_{i\neq j}J_{ij}^{{\rm H}}{\bf S}_{i}\cdot 
{\bf S}_{j}
\end{equation}
where ${\bf S}_{i}=(\sigma _{i}^{x},\sigma _{i}^{y},\sigma _{i}^{z})$ is the
Pauli matrix vector of spin $i$; (ii) the Coulomb interaction 
\begin{equation}
H_{{\rm Coul}}=\frac{1}{2}\sum_{i\neq j}\sum_{\sigma ,\sigma ^{\prime
}}J_{ij}^{{\rm C}}a_{i\sigma }^{\dagger }a_{i\sigma ^{\prime }}a_{j\sigma
^{\prime }}^{\dagger }a_{j\sigma }
\end{equation}
where $a_{i\sigma }^{\dagger }$ ($a_{i\sigma }$) is the creation
(annihilation) operator of an electron of spin $\sigma $ in Wannier orbital $
i$. $J_{ij}$ is the exchange matrix element and is given for electrons by 
\begin{equation}
J_{ij}^{{\rm C}}=-e^{2}\int d{\bf r}d{\bf r}^{\prime }\frac{w^{\ast }({\bf r}
-{\bf R}_{i})w({\bf r}^{\prime }-{\bf R}_{i})w^{\ast }({\bf r}^{\prime }- 
{\bf R}_{j})w({\bf r}-{\bf R}_{j})}{|{\bf r}-{\bf r}^{\prime }|},
\end{equation}
where ${\bf R}_{i}$ is a lattice vector and $w$ is a Wannier function \cite
{March:85vol1}. This is a rather generic form for the exchange matrix
element; in other cases $w$ would be replaced by the appropriate wave
function and the Coulomb interaction $e^{2}/|{\bf r}-{\bf r}^{\prime }|$ by
the appropriate potential. The important point to notice is that the
exchange integral depends on the overlap between the wave functions at
locations $i$ and $j$. Thus exchange effects generally decay rapidly as the
distance $|{\bf R}_{i}-{\bf R}_{j}|$ increases. An important simplification
is possible when interactions beyond nearest neighbors can be neglected
(i.e., $J_{ij}=0$ if $i$ and $j$ are not nearest neighbors) in which case
the approximation $J_{ij}\equiv J$ is often made.

In the Coulomb case the interpretation of $a_{i\sigma }^{\dagger }a_{i\sigma
^{\prime }}a_{j\sigma ^{\prime }}^{\dagger }a_{j\sigma }$ as an exchange
operator is quite clear: spin $\sigma $ is destroyed at orbital $j$ and is
created at orbital $i$, while spin $\sigma ^{\prime }$ is destroyed at
orbital $i$ and is created at orbital $j$. The net effect is that spins $
\sigma $ and $\sigma ^{\prime }$ are swapped between the electrons in
orbitals $i$ and $j$. In the Heisenberg case one can verify that the
operator ${\bf S}_{i}\cdot {\bf S}_{j}$ also implements an exchange. Let $I$
denote the identity operator, $X_{i}$ the Pauli matrix $\sigma _{i}^{x}$
operating on qubit $i$, etc. A qubit state is written as usual as a
superposition over $\sigma _{z}$ eigenstates $|0\rangle $ and $|1\rangle $.
Then, defining 
\begin{equation}
E_{ij}\equiv \frac{1}{2}\left( I+{\bf S}_{i}\cdot {\bf S}_{j}\right) =\frac{
1 }{2}\left( I+X_{i}\otimes X_{j}+Y_{i}\otimes Y_{j}+Z_{i}\otimes
Z_{j}\right) ,
\end{equation}
it is easily checked that Eq.~(\ref{eq:Eij}) is satisfied \cite{Burkard:99}.

\section{Review of Decoherence Free Subspaces}

\label{DFS-review}

We briefly recall the Hamiltonian theory of DFSs \cite
{Lidar:PRL99,Zanardi:97a}. Given is a system-bath interaction Hamiltonian 
\begin{equation}
H_{{\rm SB}}=\sum_{\lambda }F_{\lambda }\otimes B_{\lambda },
\end{equation}
where $F_{\lambda }$ and $B_{\lambda }$ are, respectively, the system and
bath operators. The decoherence free states are those, and only those states 
$\{|\psi \rangle \}$ which are simultaneous degenerate eigenvectors of all
system operators appearing in $H_{{\rm SB}}$: 
\begin{equation}
F_{\lambda }|\psi \rangle =c_{\lambda }|\psi \rangle .  \label{eq:DFS}
\end{equation}
The eigenvalues $\{c_{\lambda }\}$ do not depend on $|\psi \rangle$. The
subspace spanned by these states is a DFS, meaning that under $H_{{\rm SB}}$
the evolution in this subspace is unitary, and there is no decoherence. This
results in a passive protection against errors, to be contrasted with the
active QECC approach. Of particular interest is the case where the $
\{F_{\lambda }\}$ are collective operators, such as the total spin operators 
\begin{equation}
S_{\alpha }=\sum_{i=1}^{K}\sigma _{i}^{\alpha }\qquad \alpha =x,y,z.
\label{eq:S_a}
\end{equation}
These operators satisfy $su(2)$ commutation relations, just like the local $
\sigma _{i}^{\alpha }$ Pauli operators: 
\begin{equation}
\lbrack S_{\alpha },S_{\beta }]=2i\varepsilon _{\alpha \beta \gamma
}S_{\gamma }.
\end{equation}
This situation, referred to above as collective decoherence, arises when the
bath couples in a permutation-invariant fashion to all qubits. In this paper
we shall confine our attention to collective decoherence, and shall employ
the term DFS exclusively in this context \cite{Lidar:PRA99Pauli}. With a
system-bath interaction of the form $H_{{\rm SB}}=\sum_{\alpha }S_{\alpha
}\otimes B_{\alpha }$ (as, e.g., in the Lamb-Dicke limit of the spin-boson
model), a combinatorial calculation shows (see appendix) that the number of
encoded qubits is $\log _{2}K!/\left[ (K/2+1)!(K/2)!\right] \stackrel{
K\rightarrow \infty }{\longrightarrow }K- \frac{3}{2}\log _{2}K$. The
resulting decoherence free code thus asymptotically approaches unit
efficiency (number of encoded qubits per physical qubits), and is therefore
of significant interest. In the collective decoherence case, since the $
S_{\alpha }$ are the generators of the semisimple Lie algebra $su(2)$, the
DFS condition Eq.~(\ref{eq:DFS}) is satisfied with $c_{\alpha }=0$ \cite
{Lidar:PRL98}. This means that the decoherence free states $\{|j\rangle \}$
are $su(2)$ {\em singlets}: they are states of zero total spin, and belong
to the one-dimensional irreducible representation of $su(2)$. For example,
for $K=2$ qubits undergoing collective decoherence, there is just one
decoherence free state: $(|01\rangle -|10\rangle )/\sqrt{2} $, i.e., the
familiar singlet state of two spin 1/2 particles. For as few as $K=4$ there
are already two singlet states, spanning a full encoded decoherence free
qubit \cite{Zanardi:97c}.

\section{Decoherence Free States and Exchange with Constant Matrix Elements}

\label{const}

A simple situation arises when we can assume that $J_{ij}\equiv J/K$ for 
{\em all} $i,j$, i.e., without the restriction to nearest neighbor
interactions. This long-range Ising model is thermodynamically equivalent to
the mean-field theory of metallic ferromagnets, and there exist some
examples of metals (e.g., HoRh$_{4}$B$_{4}$) that are well described by it 
\cite{Mattis:book}. At present the relevance of such materials to quantum
computer architectures is not clear.
We also stress that in the vast majority of physical examples
exchange correlations decay exponentially fast with the distance
between particles. The case of arbitrary exchange matrix elements is
dealt with in the next section.
We consider the long-range model here
mainly for its simplicity and for the remarkable result that DFSs are
completely immune to exchange errors in this case.

We have for ${\bf S}=(S_{x},S_{y},S_{z})$ 
\begin{equation}
S^{2}={\bf S\cdot S}=3KI+2\sum_{i\neq j}X_{i}\otimes X_{j}+Y_{i}\otimes
Y_{j}+Z_{i}\otimes Z_{j},
\end{equation}
so that the exchange Hamiltonian can be rewritten as 
\begin{eqnarray}
H_{{\rm ex}} &=&\frac{J}{4K}\sum_{i\neq j}^{K}\left( I+X_{i}\otimes
X_{j}+Y_{i}\otimes Y_{j}+Z_{i}\otimes Z_{j}\right)   \nonumber \\
&=&\frac{J}{8K}\left[ \left( K^{2}-4K\right) I+S^{2}\right] .
\label{eq:Hex2}
\end{eqnarray}
Whereas the DFS condition guarantees that no decoherence is caused by the
coupling to the bath, uncontrolled unitary evolution due to the system
Hamiltonian may still pose a significant problem. This is exactly the case
in the presence of exchange errors, as described above. However, using
Eq.~(\ref{eq:Hex2}) and recalling that the DFS states have zero total
spin, we 
see that in the collective decoherence case the DFS is in fact automatically
protected against exchange errors: 
\begin{equation}
H_{{\rm ex}}|\psi \rangle =\left[ \nu I+\frac{J}{8K}S^{2}\right] |\psi
\rangle =\nu {\rm \,}|\psi \rangle ,
\end{equation}
where $|\psi \rangle $ is a DFS state and $\nu \equiv (J/K)(K^2-4K)/8$. Since the
constant $\nu $ does not depend on $\psi $, this implies that under the
unitary evolution generated by $H_{{\rm ex}}$, a DFS state accumulates an
overall, global phase $e^{i\nu {\rm \,}t}$. This phase is not measurable and
does not affect the decoherence time. Thus in the $J_{ij}\equiv J$ model a
DFS does not undergo exchange errors, and the smallest DFS ($K=4$ physical
qubits) already suffices to encode a full logical qubit.

\section{Decoherence Free States and Arbitrary Exchange Matrix Elements}

\label{arbitrary}

We now analyze the effect of arbitrary exchange errors on DFS states for
collective decoherence. We show that by concatenation with QECCs, DFSs can
be stabilized against these errors.

\subsection{Decoherence Free Subspaces are Invariant Under Exchange}

The exchange operator commutes with the total spin operators. To see this,
use the definitions of these operators in Eqs.~(\ref{eq:Eij}) and (\ref
{eq:S_a}), and let $S_{\alpha }^{ij}\equiv \left( \sum_{k\neq i,j}^{K}\sigma
_{k}^{\alpha }\right) $. Since they act on different qubits, $S_{\alpha
}^{ij}$ clearly commutes with $E_{ij}$. Now, using $\sigma ^{\alpha }\sigma
^{\beta }=\delta _{\alpha \beta }I+i\varepsilon _{\alpha \beta \gamma
}\sigma ^{\gamma }$: 
\begin{eqnarray}
S_{\alpha }E_{ij} &=&\left[ S_{\alpha }-\left( \sigma _{i}^{\alpha }+\sigma
_{j}^{\alpha }\right) \right] E_{ij}+\left( \sigma _{i}^{\alpha }+\sigma
_{j}^{\alpha }\right) E_{ij}  \nonumber \\
&=&\left( \sum_{k\neq i,j}^{K}\sigma _{k}^{\alpha }\right) E_{ij}+\frac{1}{2
}\left( \sigma _{i}^{\alpha }+\sigma _{j}^{\alpha }\right) \left(
I+\sum_{\beta =x,y,z}\sigma _{i}^{\beta }\otimes \sigma _{j}^{\beta }\right) 
\nonumber \\
&=&S_{\alpha }^{ij}E_{ij}+\sigma _{i}^{\alpha }+\sigma _{j}^{\alpha
}+\frac{i}{2}\sum_{\beta ,\gamma }\varepsilon _{\alpha \beta \gamma }\left( \sigma
_{i}^{\beta }\otimes \sigma _{j}^{\gamma }+\sigma _{i}^{\gamma }\otimes
\sigma _{j}^{\beta }\right) .
\end{eqnarray}
The last term in this expression vanishes since $\varepsilon _{\alpha \beta
\gamma }=-\varepsilon _{\alpha \gamma \beta }$ and we are summing over all $
\beta ,\gamma $ values. Thus 
\begin{equation}
S_{\alpha }E_{ij}=S_{\alpha }^{ij}E_{ij}+\sigma _{i}^{\alpha }+\sigma
_{j}^{\alpha }=E_{ij}S_{\alpha }.
\end{equation}
Now let $|\psi \rangle $ be a decoherence free state (which it is for
collective decoherence iff $S_{\alpha }|\psi \rangle =0$
\cite{Lidar:PRL98}). Since $S_{\alpha }\left( E_{ij}|\psi \rangle
\right) =E_{ij}S_{\alpha 
}|\psi \rangle =0$, it follows that $E_{ij}|\psi \rangle $ also is
decoherence free. We have thus proved:

{\it Theorem I.} Let ${\tilde {{\cal H}}}$ be a decoherence free subspace
against collective decoherence errors, and $E_{ij}$ an exchange operation on
qubits $i$ and $j$. Then $E_{ij}{\tilde {{\cal H}}}={\tilde {{\cal H}}}$.

The significance of this result is that exchange errors act as errors on the 
{\em encoded} DFS qubits, i.e., they keep decoherence free states inside the
DFS. The exact way in which these errors are manifested is a difficult
problem. Exchange operations are transpositions in the language of the
permutation group $S_K$, and are known to generate this group
\cite{Schensted}. For a given number $K$ of physical qubits the action
of the exchange operators will realize a $2^K$-dimensional reducible representation of
$S_K$. The DFS for collective decoherence on these $K$
qubits is the set of one-dimensional irreducible subspaces in the
irreducible representations (irreps) of 
$S_K$, which appear with multiplicity $\frac{K!}{(K/2+1)!(K/2)!}$ (see
appendix). For 
$K=4$ the DFS is 2-dimensional (=multiplicity of the 1D
irreps), encoding one qubit. Therefore in this case
exchange errors will act as the usual Pauli errors on a single
(encoded-) qubit. Correction of exchange errors for $K=4$ can then be
done entirely within the DFS 
by using a quantum error correcting code for single-qubit errors. This
observation naturally leads one to 
consider concatenating the DFS codewords with such a code, as done in
the concatenated code of Ref. \cite{Lidar:PRL99}. That paper 
showed that the concatenated DFS-QECC code can in fact deal with the more
general case of {\em both} errors inside the DFS (as is our case here), {\em 
and} errors that take states outside of the DFS. We investigate the
correction of exchange errors in detail for the $K=4$ case in the next
subsection. For $K>4$ qubits, the dimension of the DFS is greater than
2 (e.g., for $K=6$ it is 5), and the action of exchange errors will
correspondingly be represented by higher dimensional irreps of
$S_K$. To correct such unitary errors it will be necessary to resort
to codes for ``qu$k$its'' ($k>2$), such as stabilizer codes for
higher-dimensional sytems \cite{Gottesman:98}, or
polynomial codes \cite{Aharonov:99}. We defer the discussion of this
case to a future publication \cite{us:tbp} and focus here on the $K=4$ case.

\subsection{Effect of Exchange Errors on the Four Qubit Decoherence Free
Subspace}

Suppose that the qubits undergo collective decoherence in clusters of four
identical particles, but different clusters are independent (as they might,
e.g., in a polymer with an {\sc AAAABBBBAAAA...} type of order). Each
cluster would then support a two-dimensional DFS, accommodating a single
encoded DFS qubit. The $K=4$ physical qubits DFS states can then be written
as \cite{Zanardi:97a}

\begin{equation}
|\tilde{0}\rangle =\frac{|a\rangle -|b\rangle }{2},\quad |\tilde{1}\rangle = 
\frac{2|c\rangle -|a\rangle -|b\rangle }{2\sqrt{3}} ,
\end{equation}
where 
\begin{equation}
|a\rangle \equiv |0110\rangle +|1001\rangle ,\quad |b\rangle \equiv
|1010\rangle +|0101\rangle ,\quad |c\rangle \equiv |0011\rangle
+|1100\rangle .
\end{equation}
Note that the mutually orthogonal states $|a\rangle $, $|b\rangle $ and $
|c\rangle $ are sums of complementary states. Moreover, the four qubits play
a symmetrical role (i.e., $0$ and $1$ appear equally in all four positions
in both $|\tilde{0}\rangle $ and $|\tilde{1}\rangle $). This dictates that
exchange of qubits in symmetrical positions should have the same effect. In
other words, we expect $E_{12}$ to be indistinguishable from $E_{34}$, and
similarly for $\{E_{13},E_{24}\}$ and $\{E_{23},E_{14}\}$ (although for a
linear geometry most physical exchange mechanisms will yield $
|J_{23}|>|J_{14}|$). This expectation is born out; in the $\{|\tilde{0}
\rangle ,|\tilde{1}\rangle \}$ basis we find, using straightforward algebra,
that the six exchange operators can be written as 
\begin{eqnarray}
E_{12} &=&E_{34}=\left( 
\begin{array}{cc}
-1 & 0 \\ 
0 & 1
\end{array}
\right) =-\bar{Z}  \nonumber \\
E_{13} &=&E_{24}=\tilde{R}\left( \pi /3\right) =\frac{\sqrt{3}}{2}\bar{X}+ 
\frac{1}{2}\bar{Z}  \nonumber \\
E_{14} &=&E_{23}=\tilde{R}\left( -\pi /3\right) =-\frac{\sqrt{3}}{2}\bar{X}+ 
\frac{1}{2}\bar{Z},  \label{eq:Pauli-DFS}
\end{eqnarray}
where $\tilde{R}\left( \theta \right) =R\left( \theta \right) Z$, and $
R\left( \theta \right) =\left( 
\begin{array}{cc}
\cos \theta & -\sin \theta \\ 
\sin \theta & \cos \theta
\end{array}
\right) $. Thus\ $\tilde{R}\left( \theta \right) $ is a reflection about the 
$x$-axis followed by a counter-clockwise rotation in the $x,y$ plane. In
writing these expressions, the matrices operate on column vectors such that $
|\tilde{0}\rangle ={{{{{{\ {
{1  \choose 0}
} }}}}}}$ and $|\tilde{1}\rangle ={{{{{{\ {
{0  \choose 1}
} }}}}}},$ and $\bar{X},\bar{Z}$ are the {\em encoded} Pauli matrices, i.e.,
the Pauli matrices acting on the DFS states (and {\em not} on the physical
qubits). Thus, exchange errors act as encoded Pauli errors on the DFS states.

Using this observation, it is possible to protect DFS states against such
errors by concatenation with a QECC designed to correct single qubit errors.
The critical point is that this QECC will now correct single {\em encoded}
qubit errors. This requires an additional encoding layer to be constructed.
In particular, suppose we add such an encoding layer by using DFS qubits to
build codewords of the five-qubit ``perfect'' QECC \cite{Laflamme:96}. These
codewords have the form $|\tilde{\epsilon}_{1}\rangle |\tilde{\epsilon}
_{2}\rangle |\tilde{\epsilon}_{3}\rangle |\tilde{\epsilon}_{4}\rangle | 
\tilde{\epsilon}_{5}\rangle $, where $\epsilon =0,1$, and $j$ in $\tilde{
\epsilon}_{j}$ is now a {\em cluster} index. Since the five-qubit QECC can
correct any single qubit error, in particular it can correct the specific
errors of Eq.~(\ref{eq:Pauli-DFS}) which the encoded DFS qubits would
undergo under an exchange interaction on the physical qubits in a given
cluster. However, the error detection and correction procedure must be
carried out sufficiently fast so that exchange errors affecting multiple
blocks at a time do not occur, or else concatenation with a code that can
deal with $t>1$ independent errors is needed. The typical time scale for
exchange errors to occur is $1/2|J_{ij}|$, where $J_{ij}$ is the relevant
exchange matrix element.

This 20-qubit concatenated DFS-QECC code is precisely the one discussed in
Ref. \cite{Lidar:PRL99}, where it was shown that it offers protection
against general collective decoherence symmetry breaking perturbations. Our
present result shows that this concatenated code is stable against exchange
errors as well.

We note that it is certainly possible to find a shorter QECC than the
five-qubit one to protect against the restricted set of errors in Eq.~(\ref
{eq:Pauli-DFS}). However, such a code would not offer the full protection
against general errors that is offered by concatenation with the perfect
five-qubit code, and thus would not be as useful.

\section{Summary and Conclusions}

To conclude, in this paper we considered the effect of unitary exchange
errors between identical qubits on the protection of quantum information by
decoherence free subspaces (DFSs) defined for a qubit array. We showed that
in the important case of ideal collective decoherence (qubits are coupled
symmetrically to the bath), for which a perfectly stable DFS is obtained,
DFSs are additionally invariant to exchange errors. Thus such errors
generate rotations inside the DFS, but do not take decoherence free states
outside of the DFS. Consequently it is possible to use, without any
modification, the concatenated DFS-QECC scheme of Ref. \cite{Lidar:PRL99} in
order to protect DFSs against exchange errors, while at the same time
relaxing the constraint of ideal collective decoherence, and allowing for
symmetry breaking perturbations. This is useful for quantum memory
applications. Since exchange interactions preserve a DFS, an
interesting further question is whether they can be used {\em
constructively} in order to perform controlled logic operations inside
a DFS. We have found the answer to be positive, and 
that it is actually 
possible to perform universal computation in a fault tolerant manner
inside a DFS for collective decoherence using only two-body exchange
interactions \cite{Bacon:PRL99}.

{\it Acknowledgments}.--- This material is based upon work supported by the
U.S. Army Research Office under contract/grant number DAAG55-98-1-0371, and
in part by NSF CHE-9616615.

\appendix

\section*{Dimension of Decoherence Free Subspaces for Collective Decoherence}

In view of the fact that the total spin operators $S_{\alpha }$ satisfy
spin-1/2 commutation relations, it follows from the addition of angular
momentum that the operators $S^{2}$ and $S_{z}$ have simultaneous
eigenstates given by 
\begin{equation}
S^{2}|S,m\rangle =S(S+1)|S,m\rangle ,\qquad S_{z}|S,m\rangle =m|S,m\rangle ,
\end{equation}
where $m=-S,-S+1,...,S$ and $S=0,1,...,K/2$ (for $K$ even), $
S=1/2,3/2,...,K/2$ (for $K$ odd). The $|S,m\rangle $ states are known as
Dicke states \cite{Dicke:54,Agarwal:book}. The degeneracy of a state with
given $S$ is 
\begin{equation}
\frac{K!(2S+1)}{(K/2+S+1)!(K/2-S)!},  \label{eq:deg}
\end{equation}
which for $S=0$, i.e., the singlet states, coincides with the dimension of
the DFS for $K$ qubits undergoing collective decoherence cited in the text.

It is interesting to derive this formula from combinatorial arguments
relating to the permutation group of $K$ objects, which we will do for $S=0$. The result follows straightforwardly from the Young diagram technique. As
is well known (see, e.g., \cite{Schensted}), the singlet states of $su(2)$
belong to the rectangular Young tableaux of $K/2$ columns and $2$ rows. The
multiplicity $\lambda $ of such states is the number of ``standard
tableaux'' (tableaux containing an arrangement of numbers which increase
from left to right in a row and from top to bottom in a column), which is
also the dimension of the irreducible representation of the permutation
group corresponding to the Young diagram $\eta _{K/2,2}$ (an empty tableau)
of $K/2$ columns and $2$ rows. This number is found using the ``hook
recipe'' \cite{Schensted}, where one writes the ``hook length'' $g_{i}$ (the
sum of the number of positions to the right of box $i$, plus the number of
positions below it, plus one) of each box $i$ in the Young diagram: 
\begin{equation}
\lambda (\eta )=\frac{K!}{\prod_{i=1}^{K}g_{i}}.
\end{equation}
E.g., for $\eta _{c,2}$ the hook lengths are: 
\begin{equation}
\begin{tabular}{|c|c|c|c|c|c|}
\hline
$c+1$ & $c$ & $c-1$ & $\cdots $ & $3$ & $2$ \\ \hline
$c$ & $c-1$ & $c-2$ & $\cdots $ & $2$ & $1$ \\ \hline
\end{tabular}
\end{equation}
and one finds, with $c=K/2$: 
\begin{equation}
\lambda (\eta _{K/2,2})=\frac{K!}{(K/2+1)!(K/2)!},
\end{equation}
which is indeed the $S=0$ case of the general degeneracy formula, Eq.~(\ref
{eq:deg}).


\begin{thebibliography}{10}

\bibitem{Lo:book}  {H.K. Lo, S. Popescu and T.P. Spiller}, {\em Introduction
to Quantum Computation} (World Scientific, Singapore, 1999).

\bibitem{Ruskai:99}  {M.B. Ruskai}, {Pauli Exchange Errors in Quantum
Computation}, \uppercase{L}ANL Report No. quant-ph/9906114.

\bibitem{Shor:95Steane:96a}  (a) {P.W. Shor}, Phys. Rev. A {\bf 52}, 2493
(1995); (b) {A.M. Steane}, Phys. Rev. Lett. {\bf 77}, 793 (1996).

\bibitem{Gottesman:97Knill:97b}  (a) {D. Gottesman}, Phys. Rev. A {\bf
54}, 1862 (1997); (b) {E. Knill and R. Laflamme}, Phys. Rev. A {\bf
55}, 900 (1997).

\bibitem{Laflamme:96}  {R. Laflamme, C. Miquel, J.P. Paz and W.H. Zurek},
Phys. Rev. Lett. {\bf 77}, 198 (1996).

\bibitem{Palma:96Duan:98Zanardi:98a}  (a) {G.M. Palma, K.-A. Suominen and A.K.
Ekert}, Proc. Roy. Soc. London Ser. A {\bf 452}, 567 (1996); (b) {L.-M Duan
and G.-C. Guo}, Phys. Rev. A {\bf 57}, 737 (1998); (c) P. Zanardi,
Phys. Rev. A  {\bf 57}, 3276 (1998).

\bibitem{Zanardi:97c}  {P. Zanardi and M. Rasetti},
Phys. Rev. Lett. {\bf 79}, 3306 (1997).

\bibitem{Lidar:PRL98}  {D.A. Lidar, I.L. Chuang and K.B. Whaley}, Phys. Rev.
Lett. {\bf 81}, 2594 (1998).

\bibitem{Lidar:PRL99}  {D.A. Lidar, D.A. Bacon and K.B. Whaley}, Phys. Rev.
Lett. {\bf 82}, 4556 (1999).

\bibitem{Dicke:54} R. Dicke, Phys. Rev. {\bf 93}, 99 (1954).

\bibitem{Agarwal:book}  G. Agarwal, {\em Quantum Optics}, No.~70 in {\em 
Springer Tracts of Modern Physics} (Springer-Verlag, Berlin, 1974).

\bibitem{Knill:98Preskill:99}  (a) R. Laflamme, E. Knill and W.
Zurek, Science {\bf 279}, 342 (1998); (b) J. Preskill, in [1], p. 213.

\bibitem{Aharonov:99}  {D. Aharonov and M. Ben-Or}, {Fault-Tolerant
Quantum Computation With Constant Error Rate}, \uppercase{L}ANL Report
No. quant-ph/9906129.

\bibitem{March:85vol1}  {W. Jones and N. March}, {\em {Theoretical Solid
State Physics}} ({Dover}, {New York}, 1985), Vol.~1.

\bibitem{Burkard:99}  This observation was used by {G. Burkard, D. Loss and
D.P. DiVincenzo}, Phys. Rev. B {\bf 59}, 2070 (1999), to show that a
Heisenberg coupling enables one to implement swapping between any pair of
qubits in quantum dots.

\bibitem{Zanardi:97a}  {P. Zanardi and M. Rasetti}, Mod. Phys. Lett. B {\bf 
11}, 1085 (1997).

\bibitem{Lidar:PRA99Pauli}  DFSs may arise in other contexts as well, such
as a class that exists without any underlying spatial symmetry: {D.A. Lidar,
D.A. Bacon, J. Kempe and K.B. Whaley}, \uppercase{L}ANL Report
No. quant-ph/9908064.

\bibitem{Mattis:book}  D.C. Mattis, {\em The Theory of Magnetism II}, No. 55
in {\em Springer Series in Solid-State Sciences} (Springer-Verlag, Berlin,
1985), p. 31.

\bibitem{Schensted}  I.~V. Schensted, {\em A Course on the Application of
Group Theory to Quantum Mechanics} (NEO Press, Peaks Island, Maine, 1976).

\bibitem{Gottesman:98}  {D. Gottesman}, {Fault-Tolerant Quantum
Computation with Higher-Dimensional Systems}, \uppercase{L}ANL Report
No. quant-ph/9802007.

\bibitem{us:tbp} {J. Kempe, D. Bacon, D.A. Lidar and K.B. Whaley}, in
preparation.

\bibitem{Bacon:PRL99} {D. Bacon, J. Kempe, D.A. Lidar and
K.B. Whaley}, {Universal Fault-Tolerant Computation on
Decoherence-Free Subspaces}, \uppercase{L}ANL Report No. quant-ph/9909058.

\end{thebibliography}
\end{document}